\documentclass[runningheads]{llncs}

\usepackage[T1]{fontenc}
\usepackage{algorithm}
\usepackage{algpseudocode}
\usepackage{amsmath}
\usepackage{amssymb}
\usepackage{appendix}
\usepackage{color}
\usepackage{comment}
\usepackage{enumerate}
\usepackage{enumitem}
\usepackage{float}
\usepackage{graphicx}
\usepackage{hyperref}
\usepackage{mathrsfs}
\usepackage{mathtools}
\usepackage[normalem]{ulem}
\usepackage{xcolor}
\usepackage{xurl}

\providecommand{\N}{}

\renewcommand{\N}{{\mathbb N}}

\newcommand{\ABCD}{\texttt{ABCD}}

\newcommand{\mABCD}{\texttt{mABCD}}
\newcommand{\Rr}{\mathbf{R}}
\newcommand{\Aa}{\mathbf{A}}
\newcommand{\Bb}{\mathbf{B}}
\providecommand{\keywords}[1]
{
  \small	
  \textbf{\textit{Keywords---}} #1
}

\begin{document}

\title{Twinning Complex Networked Systems: Data-Driven Calibration of the \mABCD\ Synthetic Graph Generator}

\titlerunning{\mABCD~for Digital Twins}

\author{
    Piotr Br\'odka\inst{1,2} \and
    Micha\l{} Czuba\inst{1,2}\thanks{Corresponding author: \email{michal.czuba@pwr.edu.pl}} \and
    Bogumi\l{} Kami\'{n}ski\inst{3} \and
    \L{}ukasz Krai\'{n}ski\inst{3} \and
    Katarzyna Musial\inst{2} \and
    Pawe\l{} Pra\l{}at\inst{4} \and
    Mateusz Stolarski\inst{1}
}

\authorrunning{Piotr Br\'odka et al.}

\institute{
    \scriptsize{Dept. of Artificial Intelligence, Wroc\l{}aw University of Science and Technology, Wroc\l{}aw, Poland} \and
    \scriptsize{Data Science Institute, University of Technology Sydney, Sydney, Australia} \and
    \scriptsize{Decision Analysis and Support Unit, SGH Warsaw School of Economics, Warsaw, Poland} \and
    \scriptsize{Dept. of Mathematics, Toronto Metropolitan University, Toronto, Canada}
}

\maketitle  

\begin{abstract}
    The increasing availability of relational data has contributed to a growing reliance on network-based representations of complex systems. Over time, these models have evolved to capture more nuanced properties, such as the heterogeneity of relationships, leading to the concept of multilayer networks. However, the analysis and evaluation of methods for these structures is often hindered by the limited availability of large-scale empirical data. As a result, graph generators are commonly used as a workaround, albeit at the cost of introducing systematic biases. In this paper, we address the inverse-generator problem by inferring the configuration parameters of a multilayer network generator, \mABCD, from a real-world system. Our goal is to identify parameter settings that enable the generator to produce synthetic networks that act as digital twins of the original structure. We propose a method for estimating matching configurations and for quantifying the associated error. Our results demonstrate that this task is non-trivial, as strong interdependencies between configuration parameters weaken independent estimation and instead favour a joint-prediction approach.
\end{abstract}


%%%%%%%%%%%%%%%%%%%%%%%%%%%%%%%%%%%%%%%%%%%%%%%%%%%%%%%%%%%
\section{Introduction}\label{sec:intro} 
%%%%%%%%%%%%%%%%%%%%%%%%%%%%%%%%%%%%%%%%%%%%%%%%%%%%%%%%%%%

Complex networked systems are typically represented by graphs constructed from relational data. This representation facilitates the encapsulation of topological characteristics, such as relationship structures, connectivity patterns, or node centralities, which are not readily apparent at first glance~\cite{easley2010networks}.

To accurately characterise dynamics in complex networked systems, it is necessary to adopt an approach that differentiates between the various types of relationships present. A natural way to achieve this is through the use of \textit{multilayer networks}~\cite{kivela2014multilayer}, a system defined as a collection of $\ell$ layers, where each layer represents a distinct mode of interaction encoded by its corresponding graph. The resulting dynamics are governed not only by intra-layer edges, describing interactions within a given relationship type, but also by the presence of the system's actors across different layers, through which they serve as inter-layer bridges. This structure enables modelling of complex spreading mechanisms that are obscured in monoplex representations~\cite{dedomenico2016physics}.

Understanding the topology of a system and the interplay between the relationships it contains is critical for designing effective intervention policies. For instance, in simplified models of human interactions, immunisation strategies (such as node removal or fact-checking) typically target high-degree hubs~\cite{tambuscio2015fact}. However, in multilayer systems, the most influential actors are often not the hubs within individual layers, but rather multiplex bridges, that is, nodes with moderate intra-layer connectivity but high ``participation coefficients'' across layers. These actors facilitate the spillover of misinformation from fringe, weakly moderated communities (e.g., \textit{4chan}) into mainstream discourse (e.g., \textit{X} or \textit{Facebook}). Consequently, observing the system through a multilayer lens reveals that resilience in one layer can mask catastrophic fragility in the coupled system. A misinformation cascade that appears sub-critical (dying out) on one platform may be sustained by activity on another, creating a hysteresis loop where the false belief becomes endemic despite platform-specific moderation efforts.

While empirical analysis of real-world datasets is foundational, it faces severe limitations when dissecting the complex dynamics of multilayer systems. Real-world multilayer data is often sparse, proprietary, or plagued by missing links, and crucially, it represents only a single, static realisation of a dynamic process. One cannot ``rewind'' the Internet to see how a misinformation campaign would have unfolded if the user base were 50\% larger or if the moderation algorithm were different. To overcome these constraints, researchers must turn to synthetic network generation, that is, creating artificial yet statistically realistic graphs that serve as controllable laboratories for simulation. The primary utility of these synthetic environments lies in their flexibility. By employing generative random graph models, one can systematically vary topological parameters to isolate their effects on dynamical processes.

This approach culminates in the concept of the ``\emph{digital twin}''~\cite{wen2024dtcns}. Digital twins for complex networked systems provide a dynamic virtual counterpart to real-world networks, enabling continuous monitoring, simulation, and prediction of system behaviour. Their defining strength lies in real-time data exchange between the physical system and its digital representation, allowing the model to evolve as the network evolves. By integrating simulation, optimisation, data analytics, and machine learning, digital twins capture both the structure and the dynamics of interconnected systems such as social networks, cyber-physical systems, and blockchain-based architectures. This makes them uniquely suited to exploring scenarios that have never occurred, detecting anomalies, and supporting informed decision-making in complex, distributed environments. Even on the simplest level of modelling, a digital twin is not merely a random graph, but a synthetic replica carefully calibrated to match the specific statistical properties of a target real-world system (such as its degree distribution and distribution of community sizes, etc.). Therefore, the modelling paradigm they convey makes them a powerful approach for understanding and managing complex systems.

While the utility of ``digital twins'' is clear, constructing one that faithfully mimics a specific real-world system remains a formidable mathematical challenge. This is fundamentally an inverse problem: given an observed set of real-world statistics, one must infer the precise combination of generative parameters that produces a matching graph. In multilayer systems, this difficulty is compounded by the curse of dimensionality. A model rich enough to capture the nuance of real systems, accounting for intra-layer topology and inter-layer interdependence, inevitably possesses a vast parameter space. Developing robust, algorithmic frameworks to automatically calibrate these high-dimensional synthetic models to empirical data is not just a technical refinement; it is a necessary frontier for making network science predictive rather than merely descriptive.

In this paper, we aim to shed light on this problem by proposing a modular approach for inferring the parameters of the \mABCD\ generator from an observed real-world network. In particular, we investigate whether decomposing the task into a sequence of analytical and optimisation sub-problems constitutes a viable strategy. Our results indicate that the proposed approach establishes a strong baseline that could be difficult to surpass. Nevertheless, the experiments suggest that jointly predicting all configuration parameters within a single optimisation procedure may yield superior performance, which delineates an interesting yet challenging new research direction.

The remainder of the paper is organised as follows. Sec.~\ref{sec:preliminaries} briefly describes the fundamental concepts underpinning the work; in particular, operating principles of the \mABCD\ model. Sec.~\ref{sec:extracting_parameters} presents the parameter estimation framework and introduces the discrepancy measures used to quantify approximation error. Sec.~\ref{sec:experiments} discusses the experimental results, while Sec.~\ref{sec:conclusions} concludes the paper.

%%%%%%%%%%%%%%%%%%%%%%%%%%%%%%%%%%%%%%%%%%%%%%%%%%%%%%%%%%%
\section{Preliminaries\label{sec:preliminaries}}
%%%%%%%%%%%%%%%%%%%%%%%%%%%%%%%%%%%%%%%%%%%%%%%%%%%%%%%%%%%

To address the problem studied in this paper, we first introduce the fundamental concepts. We briefly review multilayer networks, outline the general concept for the configuration inference task, and describe the synthetic graph generator employed.

\subsection{Multilayer networks}

As stated in the introduction, heterogeneity of complex networked systems is a crucial property which cannot be neglected. Therefore, the problem tackled in this work aims to take into account this property by utilising multilayer networks. Following~\cite{kivela2014multilayer}, we formalise the concept of a multilayer network below. Nevertheless, before we do this, we clarify that for a given $n \in \N = \{1, 2, \ldots \}$, $[n]$ is used to denote the set consisting of the first $n$ natural numbers, that is, $[n] = \{1, 2, \ldots, n\}$. Furthermore, for a matrix $\mathbf{X}$, we denote its $(i,j)$-th entry by $\mathbf{X}_{ij}$.

\begin{definition}[Multilayer network]
    \label{def:multilayer_net}
    A multilayer network can be described as a quadruple $G = ([n],[\ell],V,E)$, where $[n]$ is a set of $n$ actors, $[\ell]$ is a set of layers, $V \subseteq [n] \times [\ell]$ is a set of nodes, $E \subseteq  \bigcup_{i \in [\ell]} [V_i \times V_i]$ is a set of edges.
\end{definition}

When analysing  Def.~\ref{def:multilayer_net}, one can note that it is in line with the understanding of multilayer networks as collections of graphs, where each layer $G_i = (V_i, E_i)$ is spanned on a subset of actors existing in the system ($v_i$) within the relationship (layer) $i \in [\ell]$. For instance, in a system comprising interactions on various social platforms, $[n]$ would denote humans active on the Internet, $[\ell]$ social platforms, $V$ would be a set of all accounts in the systems, i.e.\ nodes, where node $v = (a,\ell_i) \in V$ represents an actor $a$ in layer ${\ell}_i$, e.g., a \textit{LinkedIn} profile of a citizen John Smith. On the other hand, the network could be decomposed into $\ell$ interdependent graphs from different social systems.

\subsection{Complex Networked System Analysis with Twinning Approach}

Utilising the example introduced in Sec.~\ref{sec:intro}, a conceptual framework guiding the problem is presented here. Suppose one is to evaluate several ``what-if'' scenarios that cover interventions in a social network to design effective strategies to control the spread of misinformation, given an initial dataset of ground-truth interactions on the platform. 

\begin{figure}[ht!]
	\centering
	\includegraphics[width=\linewidth]{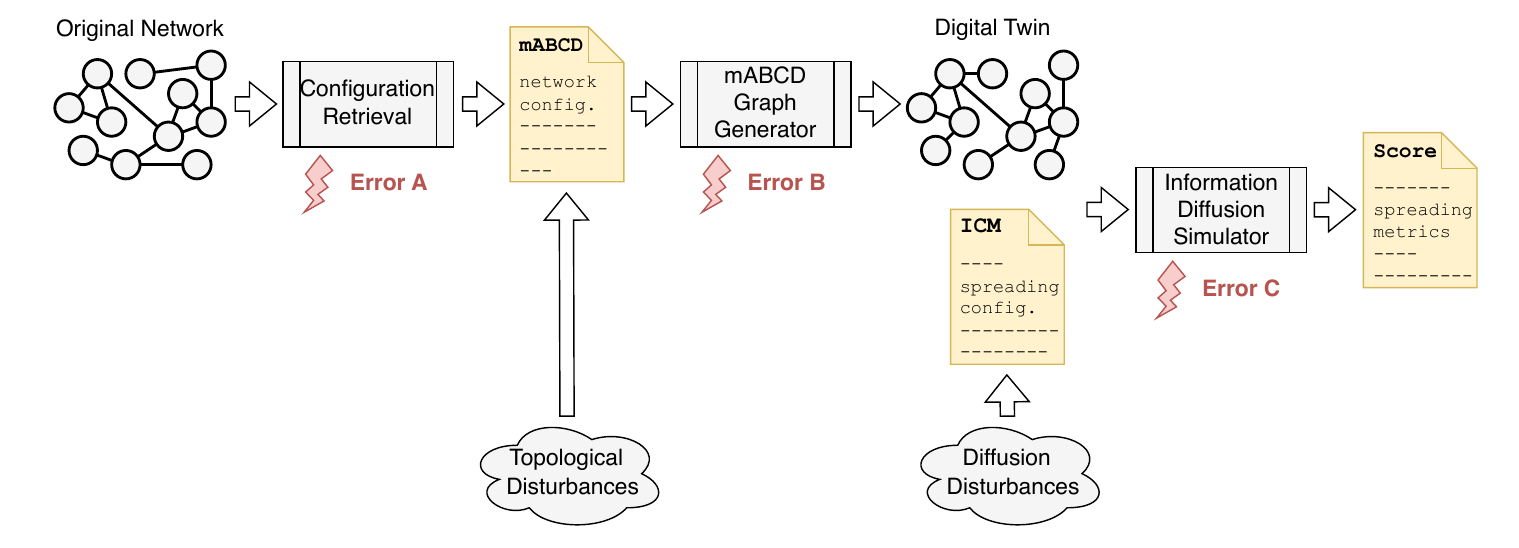}
    \vspace{-2em}
    \caption{Spread control pipeline utilising the configuration retrieval approach to create adjusted twins of the evaluated system. The process begins by determining the parameters of \mABCD\ that match the analysed system. Its digital twin is then generated and evaluated under the appropriate dynamics. Note that the pipeline's efficacy depends on addressing inherent errors.}
    \label{fig:pipeline}
\end{figure}

On the grounds presented in Sec.~\ref{sec:intro}, a twin-based framework can be introduced, as shown in Fig.~\ref{fig:pipeline}. It assumes that the analysed real-world multirelational system is encoded into a configuration file for a synthetic graph generator, which can then be used for further modelling. By altering selected parameters of the generator, one can explore a variety of what-if scenarios and construct adjusted ``digital twins'' that reflect possible modifications of the original system. These can be evaluated under the assumed spreading regime, allowing an assessment of which topology is optimal from the perspective of a given influence control task (e.g., maximisation). For instance, one can rigorously test hypotheses by scaling the network size to observe asymptotic behaviours, increasing edge density to identify percolation thresholds, or introducing additional layers to simulate the emergence of new communication channels. This capability is essential for establishing causal links between network structure and process outcomes, something that is impossible with observational data alone.

The most important factor affecting the validity of the approach involves three types of errors. The first one arises in the \textit{configuration retrieval} process. For models allowing rich modelling space (like \mABCD), this task is non-trivial, as some parameters cannot be directly inferred from the network and need to be estimated (e.g., a community breakdown). The second type of error relates to imperfections of the synthetic graph generator, as the produced network may deviate from the specified configuration. Finally, the third type of error arises in the simulation of information diffusion, since spreading models remain only approximations of real-world processes. Addressing these errors is essential to ensure the pipeline's correctness. Given the size of the task, we aim to address the first of the presented errors in this work.

Configuration retrieval can be carried out in different ways depending on needs and availabilities. The most natural approach is to estimate each parameter independently by designing appropriate algorithms and metrics to quantify estimation quality. This, however, is not the only method. Another, perhaps complementary, natural way of measuring the ``distance'' between the original network $G$ and its produced twin $\hat{G}$ would be to embed both graphs in the same high-dimensional space and use the distance between the two representations to measure the quality (\emph{Open Problem~1}). Given the recent advancements in machine learning, this approach also feels natural, and some fast algorithms were recently introduced~\cite{dehghan2025network}. Nevertheless, approaching the problem without trying the more straightforward methods deprives us of a reliable baseline. Therefore, this paper tackles the problem from the former position.

\subsection{The Synthetic Network Generators --- \mABCD\label{subsec:mabcd}}

Synthetic graph generators with explicit community structure play a central role in the development and evaluation of analysis methods for complex networks, particularly in settings where large-scale empirical datasets with reliably annotated communities are scarce. To address this limitation, several benchmark models have been proposed to generate artificial networks that resemble real-world systems while retaining full control over their structural properties. A prominent example is the \texttt{LFR} model~\cite{lancichinetti2008benchmark} and its extensions~\cite{brodka2016method}, which allows for heterogeneous degree distributions and community sizes and has become a standard tool for evaluating community detection algorithms.

More recently, the \texttt{A}rtificial \texttt{B}enchmark for \texttt{C}ommunity \texttt{D}etection (\ABCD) model~\cite{kaminski2021artificial} has been introduced as a scalable~\cite{kaminski2022properties} and theoretically tractable alternative to \texttt{LFR}. Undirected variants of \ABCD\ generate networks with comparable structural properties, while offering improved computational performance and greater flexibility in interpolating between well-defined community structure and random graphs~\cite{kaminski2022properties}. Owing to these properties, \ABCD\ has been extensively analysed from a theoretical perspective, including studies of its modularity behaviour~\cite{kaminski2022modularity} and self-similarity properties~\cite{self-similarityABCD}. The modular design of the model has further enabled a range of extensions, including outliers~\cite{kaminski2023artificial}, overlapping communities~\cite{barrett2025artificial}, hypergraphs~\cite{kaminski2023hypergraph}, and multilayer networks~\cite{krainski2025mabcd}. The multilayer extension, \mABCD, generalises the \ABCD\ framework to systems in which multiple interaction layers jointly shape network dynamics. Rather than aiming to exactly reproduce a specific empirical network, \mABCD\ is intended to generate ensembles of graphs that preserve key structural characteristics of a target system, and can therefore be interpreted as coarse-grained digital twins of complex networked systems.

In this work, we adopt \mABCD\ as the synthetic generator underlying our experimental analysis. Therefore, to focus on the aspects most relevant to the configuration inference problem considered here, a brief review of its core principles is necessary. In principle, the \mABCD\ network generation comprises of six subsequent phases. The process is governed by global and local (independent for each layer) parameters presented briefly in Tab.~\ref{tab:mabcd_parameters}. For a detailed description, please refer to~\cite{krainski2025mabcd}.

\begin{table}[ht]
    \caption{Global and local parameters of \mABCD\ denoted together as $\Theta$.}
    \centering
    \begin{tabular}{p{4.5em}p{9em}p{22em}}
    \hline
    Parameter & Range & Description \\
    \hline
    \multicolumn{3}{l}{\emph{Global parameters}} \\
    $n$ & $\N$ & Number of actors \\
    $\ell$ & $\N$ & Number of layers \\
    $d$ & $\N$ & Dimension of the biscuit-like reference layer \\
    $\Rr$ & $[0,1]^{\ell \times \ell}$ & Inter-layer edge correlation matrix \\
    \hline
    \multicolumn{3}{l}{\emph{Local parameters (independent for each $i$ layer)}} \\
    $q_i$ & $(0,1]$ & Fraction of active actors \\
    $\tau_i$ & $[-1,1]$ & Correlation coef. between degrees and actors' labels \\
    $r_i$ & $[0,1]$ & Correlation strength between communities and the reference layer \\
    $\gamma_i$ & $(2,3)$ & Power-law degree distr.\ with exponent $\gamma_i$ \\
    $\delta_i$ & $\N$ & Min.\ degree \\
    $\Delta_i$ & $\N \; (1 \le \delta_i \le \Delta_i < n)$ & Max.\ degree \\
    $\beta_i$ & $(1,2)$ & Power-law community size distr.\ with exponent $\beta_i$ \\
    $s_i$ & $\N$ & Min.\ community size \\
    $S_i$ & $\N \; (\delta < s_i \le S_i \le n)$ & Max.\ community size \\
    $\xi_i$ & $(0,1)$ & Level of noise \\
    \hline
    \end{tabular}
    \label{tab:mabcd_parameters}
\end{table}

The first five steps are carried out independently for each of the $\ell$ layers, while the sixth one binds the generated graphs into a single multilayer network. In the first phase, the so-called active nodes (i.e., those that will not be isolated) are determined. Each active node is endowed with its degree in the second step. During the third step, communities (understood as in~\cite{clauset2004greedy}) are created and populated with nodes. This is done with the help of the reference latent layer, which is implicitly generated during the network generation process and discarded afterwards. It serves as a proxy for actors' features, which typically shape connections in complex networked systems. The layer can be visualised as a $d$-dimensional biscuit, with raisins representing actors: the closer two raisins are, the more likely the corresponding actors are to be assigned to the same community. Phase four focuses on connecting nodes within the communities as well as establishing inter-community (i.e., background) links. These graphs are subsequently simplified to remove possible self-loops and multiedges while preserving the intended structural properties. The final phase performs a series of edge rewirings to achieve the targeted edge correlations between each pair of layers.

%%%%%%%%%%%%%%%%%%%%%%%%%%%%%%%%%%%%%%%%%%%%%%%%%%%%%%%%%%%
\section{Proposed \mABCD\ Configuration Retrieval Method\label{sec:extracting_parameters}}
%%%%%%%%%%%%%%%%%%%%%%%%%%%%%%%%%%%%%%%%%%%%%%%%%%%%%%%%%%%

Suppose that we are given a multilayer network following Def.~\ref{def:multilayer_net}. Our goal is to generate a digital twin, a multilayer network $\hat{G}$ generated by \mABCD\ with associated graphs $\hat{G}_i = (\hat{V}_i, \hat{E}_i)$ for $i \in [\ell]$, that mimics the original network $G$ as best as possible. In order to do it, one needs to appropriately select the parameters $\Theta$ of \mABCD\ and then measure the ``success'', i.e., how close $G$ to $\hat{G}$ is. The success is measured by computing a \emph{divergence score} $\mathcal{D} ( \hat{G}, G )$; the smaller the score, the better the fit is obtained.

As already noted, \mABCD\ consists of several parameters, but not all of them are used directly to model the target network's properties. Some should be treated rather like hyperparameters, by manipulating which the accuracy of transforming a given configuration setup into the final network can be controlled. These were intentionally omitted in Sec.~\ref{subsec:mabcd}, as they are out of our interest due to a loose connection with the structural properties of the produced network, but we acknowledge them to show the complexity of the problem. In the proposed approach, we fix them to the values recommended in~\cite{krainski2025mabcd}.

The remaining \mABCD\ parameters (Tab.~\ref{tab:mabcd_parameters}) explicitly provide expected structural properties of the network and can be grouped by their proximity. Based on the observation, we propose a method that can independently predict each parameter group by leveraging an algorithmic approach. The only exemption is $r$, which cannot be feasibly estimated as it employs the latent layer used to model actor features --- neither returned by the model nor available in real-world networks. That is why a solution-search-browsing technique was employed, i.e., the Bayesian optimisation.

\medskip

Parameters such as $n$, $\ell$, $\delta_i$, and $\Delta_i$ are trivial to extract from $G$. Similarly, the fraction of active actors is easy to obtain: $q_i = |V_i| / | \bigcup_{i \in [\ell]} V_i|$. 

\subsubsection{Dimension of reference latent layer ($d$).\label{subsec:cookie}}

Communities in \mABCD\ are independently generated in each layer, but the desired correlations are obtained via a hidden, low-dimensional geometric reference layer. Nodes close to each other in the reference layer end up in the same community with a larger probability. In particular, the dimension of a reference layer $d$ affects many properties of the generated network (see Sec.~4 in \cite{krainski2025mabcd}). However, it seems not so easy to guess the right value so we leave it as a parameter that eventually can be tuned for the best outcome, that is, the smallest divergence score $\mathcal{D} ( \hat{G}, G )$. Understanding the influence of this hyperparameter and selecting the optimal value is left as an open problem (\emph{Open Problem~2}).

This geometric approach to community structure is justified by the fact that latent geometric spaces are widely believed to shape complex networks (e.g., social media networks shaped by users' opinions, education, knowledge, interests, etc.). These latent spaces have been successfully employed for many years to model and explain network properties such as self-similarity~\cite{serrano2008self}, homophily, and aversion~\cite{henry2011emergence}. For further references, we direct the reader to the survey~\cite{boguna2021network} or the book~\cite{serrano2021shortest}.

\subsubsection{Inter-layer edge correlation ($\Rr$).}

To measure correlations between edges in different layers, we define $\Rr$, an $\ell\times\ell$ matrix in which elements $\Rr_{i,j} \in [0,1]$ ($i,j \in [\ell]$) capture correlation between edges present in layers $i$ and $j$. For any $i,j \in [\ell]$, let $E_i^j$ be the set of edges that are present in layer $i$, involving actors that are also active in layer $j$. Entries in $\Rr$ are computed using the following formula:
$$
    \Rr_{i,j} = \frac{| E_i^j \cap E_j^i |}{\min\{|E_i^j|,|E_j^i|\}}.
$$
If $\min\{|E_i^j|,|E_j^i|\}=0$, then we leave $\Rr_{i,j}$ undefined.

Note that $\Rr_{i,i}=1$ for any $i \in [\ell]$ and $\Rr_{i,j}=\Rr_{j,i}$ for $1 \le i < j \le \ell$. The maximum value of $1$ is attained when edges in one of the layers form a subset of edges in the other layer. The minimum value of $0$ is attained when the two sets of edges in the corresponding layers are completely disjoint. 

\medskip

Extracting $\Rr$ from the original network $G$ is straightforward. The \mABCD\ model aims to produce a synthetic network with the desired edge correlation matrix, but this is not guaranteed, and some small error always occurs. Let $\hat{\Rr}$ be the corresponding edge correlation matrix for the obtained digital twin $\hat{G}$. The divergence score is simply the normalised to $[0,1]$ Frobenius norm between the two matrices:
$$
\mathcal{D}_{\Rr} ( \hat{G}, G ) = \frac {\lVert {\Rr}-\hat{\Rr} \rVert_F}{\sqrt{\ell (\ell-1)}} 
= \sqrt { \frac {1}{\ell (\ell-1)} \sum_{i,j \in [\ell]} (\Rr_{ij} - \hat{\Rr}_{ij})^2 } 
= \sqrt { \frac {1}{\binom{\ell}{2}} \sum_{1 \le i < j \le \ell} (\Rr_{ij} - \hat{\Rr}_{ij})^2 }.
$$

\subsubsection{Correlation coefficient between degrees and labels ($\tau_i$).}

This family of parameters models correlations between sequences of node degrees that are active in two different layers. The \mABCD\ model assumes that the labels of actors and nodes representing them are natural numbers. The parameter $\tau_i \in [-1,1]$ models the correlation between these labels and the corresponding degree sequence in layer $i$. 

To estimate $\tau_i$, which relies on the ordering of actor labels, we first sort actors with respect to their total degree (sum of degrees over all layers), breaking ties randomly if needed; that is, the actor with the highest total degree is assigned label $n$, the highest possible label. Then, for any $i \in [\ell]$, we compute the Kendall rank correlation coefficient $\tau_i$ between these labels and the corresponding degree sequence in layer $i$. Moreover, only nodes with a positive degree are taken into account to obtain the correlation value in order to reduce the noise; i.e., inactive nodes are ignored as they should be. 

To measure the quality of the fit, one can compute an $\ell\times\ell$ matrix $\Aa$, where $\Aa_{ij} \in [-1,1]$ is the Kendall rank correlation coefficient between layers $i$ and $j$ for the real network $G$. Clearly, only nodes that are active in both layers (i.e., nodes in $V_i \cap V_j$) are considered. Once a digital twin is generated, matrix $\hat{\Aa}$ is computed analogously, but this time for the twin $\hat{G}$. As before, the divergence score is simply the (normalised) Frobenius norm between the two matrices:
$$
\mathcal{D}_{\tau} ( \hat{G}, G ) = \frac {\lVert \Aa - \hat{\Aa} \rVert_F}{\sqrt{4 \ell (\ell-1)}} = \sqrt { \frac {1}{4 \binom{\ell}{2}} \sum_{1 \le i < j \le \ell} (\Aa_{ij} - \hat{\Aa}_{ij})^2 }.
$$

\subsubsection{Correlation strength between communities and the reference layer ($r_i$).}

This family of parameters, which models correlations between communities formed by nodes that are active in two different layers, appears to be the most challenging to retrieve. That results from inaccessibility of the latent biscuit-like reference layer (whence the communities are sampled and whose dimension is modelled by $d$) used during network generation. However, once a digital twin is generated, the success of estimating $r$ can be assessed by computing the corresponding divergence score $\mathcal{D}_{r}(\hat{G}, G)$, which measures a feature indirectly captured by $r$ --- the cross-layer alignment of communities between $G$ and $\hat{G}$. By efficiently searching the parameter space to identify the value $\hat{r}$ that minimises $\mathcal{D}_{r}(\hat{G}, G)$, one can attempt to fit this parameter.

Let us then concentrate on the divergence score. For each layer $i \in [\ell]$ in the real network $G$, we independently run some stable clustering algorithm (in experiments, the Greedy Modularity Optimisation~\cite{clauset2004greedy}) to get a partition $\mathcal{P}_i$ that identifies its community structure. Then, we compute an $\ell\times\ell$ matrix $\Bb$, where $\Bb_{ij} \in [0,1]$ is the adjusted mutual information (AMI) between partitions $\mathcal{P}_i$ and $\mathcal{P}_j$ induced by nodes that are active in both layers, that is, partitions induced by $V_i \cap V_j$. Matrix $\hat{\Bb}$ is computed analogously but for the twin $\hat{G}$ and the divergence score is defined as follows:
$$
\mathcal{D}_{r} ( \hat{G}, G ) = \frac {\lVert \Bb - \hat{\Bb} \rVert_F}{\sqrt{\ell (\ell-1)}} = \sqrt { \frac {1}{\binom{\ell}{2}} \sum_{1 \le i < j \le \ell} (\Bb_{ij} - \hat{\Bb}_{ij})^2 }.
$$

The proposed method for predicting $r$ is based on access to the previously estimated parameters of the \mABCD\ configuration ($\Theta$) and the $\Bb$ matrix of the original network. The objective function generates $n$ candidate twins from the configuration $\Theta^{(r)}$ corresponding to the evaluated value of $r$. It then computes a $\hat{\Bb}$ matrix for each twin and returns the divergence score $\mathcal{D}_{r}$ between $\Bb$ and the mean matrix obtained from the collection of $\hat{\Bb}$ matrices, denoted by $\bar{\Bb}$. This value is subsequently minimised during the Bayesian optimisation process to obtain the best-matching vector $\hat{r}$:
$$
\hat{r} = \arg \min_{r \in [0,1]^\ell} \mathcal{D}_r (\Bb ,\, \bar{\Bb}): \ \bar{\Bb} = \frac{1}{n} \sum_{k=1}^{n} \hat{\Bb}^{(k)}(\Theta^{(r)})
$$

\begin{figure}[ht!]
    \centering
    \vspace{-1em}
    \includegraphics[scale=0.55]{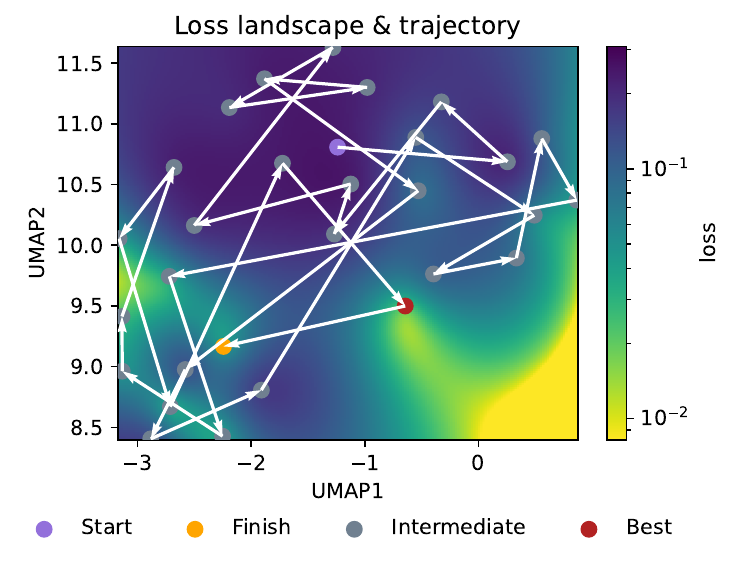}
    \caption{A typical optimisation trajectory of $r$ projected on $\mathbb{R}^2$ space obtained during experiments. Note the preserved trade-off between exploration and exploitation, by which the employed method avoids getting stuck in local minima.} 
    \label{fig:trajectory}
\end{figure}

By sequentially evaluating the objective function at values of $r$ selected based on previous evaluations, the method seeks to identify the best-matching value of~$r$. The implementation employed, provided by the \texttt{skopt}~\cite{skopt2026minimize} library, accounts for the non-deterministic behaviour of \mABCD\ by incorporating Gaussian noise into the surrogate model of the objective. A typical optimisation trajectory for fitting $r$ is shown in Fig.~\ref{fig:trajectory}.

\subsubsection{Degree distributions ($\gamma_i, \delta_i, \Delta_i$).}

This family of parameters models the degree sequences in the corresponding layers. Among them, only the retrieval of $\gamma_i$ is non-trivial, as it involves the general challenge of fitting an observed distribution to a specific model. To estimate this parameter, we employ the \texttt{powerlaw} library~\cite{alstott2014powerlaw}, implicitly assuming that the degree distribution follows a power-law.

The \mABCD\ model generates, by design, power-law sequences with parameters provided as the input. Hence, the only discrepancy between the original network $G$ and its synthetic  twin $\hat{G}$ comes from fitting the power-law distribution to the observed degree sequence of the real network. To measure the quality of the fit, we employ the Kolmogorov--Smirnov test. It is a nonparametric statistical test used to determine if a sample comes from a specific distribution (one-sample) or if two samples come from the same distribution (two-sample) by measuring the maximum difference between their cumulative distribution functions. This leads to the following divergence score:
$$
\mathcal{D}_{\gamma} ( \hat{G}, G ) = \frac {1}{\ell} \sum_{i \in [\ell]} \max_{k\in[\delta_i,\Delta_i]}\left| \frac{N_i(k)}{N_i(\delta_i)} -
\frac{\int_{k}^{\infty}x^{-\gamma_i} dx}{\int_{\delta_i}^{\infty}x^{-\gamma_i} dx} \right|,
$$
where $N_i(k)$ is the number of nodes of degree at least $k$ in layer $i$ (as a result, as usual, we ignore inactive nodes in this layer). 

\medskip

For simplicity, in this paper we assume that the degree sequences within layers of a real multilayered network $G$ (i.e., specific to nodes rather than actors) approximately follow a power law. But not all networks have degree sequences of this nature. To solve this issue, the \mABCD\ model is flexible and allows the degree sequences to be directly injected into the model. Since \mABCD\ is using the classical configuration model to generate edges in each layer, with this more advanced option, the degree sequence of the digital twin $\hat{G}$ would match \emph{exactly} the original network $G$. We leave an implementation of this extension into our framework and an investigation of its consequences on the quality of the digital twin for the future (\emph{Open Problem 3}).

\subsubsection{Distributions of community sizes ($\beta_i, s_i, S_i$).}

The next family of parameters governs distributions of community sizes in the corresponding layers. Recall that we already extracted the community structure in each layer $i \in [\ell]$ of the real network $G$ by running some stable clustering algorithm, partitions $\mathcal{P}_i$. We may now compute the corresponding sequences of community sizes and extract the values of $s_i$ and $S_i$ from these sequences. Following the same procedure as for dealing with the degree distributions, one can estimate the parameters $\beta_i$ and measure the success by computing the corresponding divergence score $\mathcal{D}_{\beta} ( \hat{G}, G )$. As before, we leave it for future work to investigate how much one gains if the sequence of community sizes is directly injected into the \mABCD\ model (\emph{Open Problem~4}).

\subsubsection{Level of noise ($\xi_i$).}

Finally, to estimate the layer-wise noise level between communities, we again analyse the community structure $\mathcal{P}_i$ of each network layer $i \in [\ell]$. For each layer~$i$ of the original multilayered network $G$, the parameter $\xi_i$ is extracted by simply looking at the fraction of inter-community edges (edges connecting nodes from different parts of the partition $\mathcal{P}_i$) relative to the total number of edges in the layer~$i$. The \mABCD\ model aims to preserve this property, but some very small error could be introduced. More importantly, the model preserves the level of noise between the ground-truth partition, which might not be exactly the same as the partition found by the clustering algorithm (especially for noisy graphs). Hence, once the twin $\hat{G}$ is generated, we extract the associated parameters $\hat{\xi}_i$ and compute the divergence score:
$$
\mathcal{D}_{\xi} ( \hat{G}, G ) = \sqrt { \frac {1}{\ell} \sum_{i \in [\ell]} (\xi_i - \hat{\xi}_i)^2 }.
$$

\subsubsection{Cumulative Divergence Score.} 

The parameters of \mABCD\ affect some important properties of multilayered networks; hence, it is desired for the digital twin $\hat{G}$ to match these properties as best as possible. For each parameter of the model, we introduced above some natural ways to measure how well the corresponding property is preserved via their divergence scores. To estimate an overall quality of the digital twin, one needs to somehow combine these scores. Their definitions guarantee that each of the scores is in $[0,1]$, but they are not comparable. Hence, before we develop a framework for generating good digital twins, we need to come up with a good measure of quality that combines all aspects together (\emph{Open Problem~5}).

%%%%%%%%%%%%%%%%%%%%%%%%%%%%%%%%%%%%%%%%%%%%%%%%%%%%%%%%%%%
\section{Experiments\label{sec:experiments}}
%%%%%%%%%%%%%%%%%%%%%%%%%%%%%%%%%%%%%%%%%%%%%%%%%%%%%%%%%%%

To test the proposed configuration retrieval method, two experiments using the framework proposed in Sec.~\ref{sec:extracting_parameters} were performed. In both of them a real-world multilayer network \textit{Freebase}~\cite{wang2021self}, representing cooperation in the movie industry, was employed. The \textit{Freebase} network captures relations between movies and people involved in their production (\textbf{A}ctors, \textbf{W}riters, \textbf{D}irectors). No explicit inter-layer edges are defined (which aligns with Def.~\ref{def:multilayer_net}); layers correspond to meta-path relations (\textbf{MAM}, \textbf{MDM}, \textbf{MWM}) over the same movie nodes. The dataset is used here only as an illustrative example, and its semantics are not considered further. The network has 3,492 actors; the numbers of nodes per layer are 3,479, 2,091, and 1,865, with 129,097, 7,099, and 5,948 edges, respectively. The mean degree is 75.41 and the average closeness centrality is 0.1.

\subsection{Experiment 1 -- Dimensionality of the Reference Layer \textit{d}}

The objective of the first experiment was to evaluate the influence of the dimension of the reference layer ($d$) on the divergence scores between the original network and its produced twins. As such, the estimation procedure was performed independently for four fixed values of $d$, i.e. $d\in \{ 1,2,4,8 \}$. Note that it directly affects the Bayesian optimisation component of the proposed method that is used to estimate $r$.

\begin{figure}[ht!]
    \centering
    \vspace{-2em}
    \includegraphics[scale=0.51]{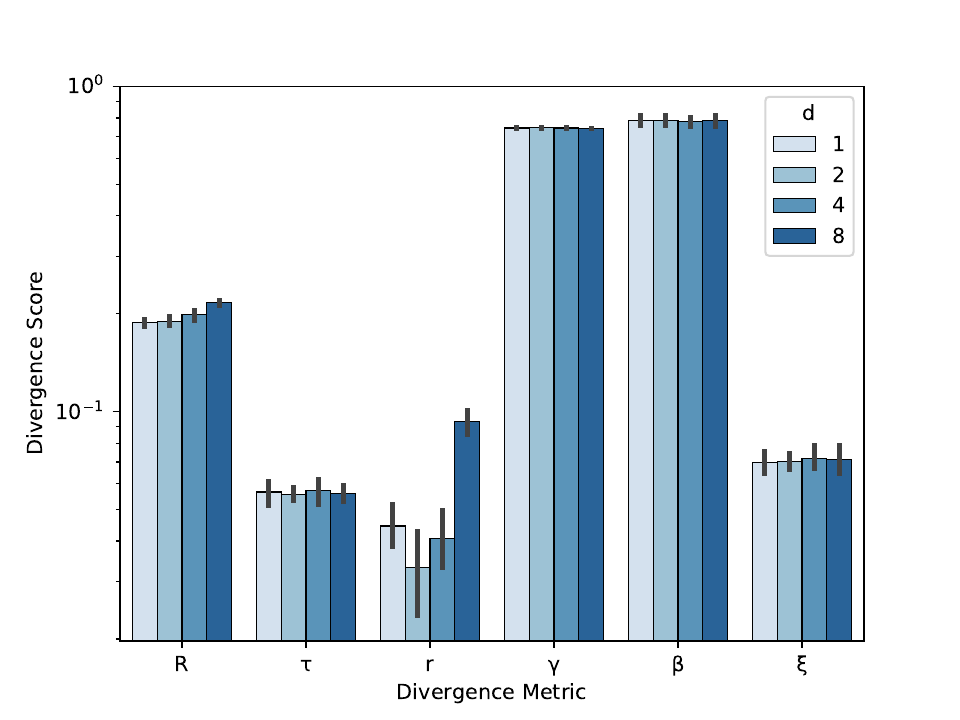}
    \caption{Divergence scores for configuration retrieval with fixed $d = 2^k$ dimension, $k\in \{ 0, 1, 2, 3 \}$; estimation with tuned $r$ and the optimisation loss $\mathcal{D}_{r} ( \hat{G}, G )$.} 
    \label{fig:experiment_1}
\end{figure}

Fig.~\ref{fig:experiment_1} presents a comparison of the average values of the corresponding divergence metrics across the tested $d$. The scores associated with $\gamma$, $\beta$, $\xi$, and $\tau$ remain virtually unchanged, as expected, since these parameters were fixed during the optimisation of $r$. Their stability only confirms the reproducibility of the proposed approach. In contrast, a pronounced effect of $d$ is observed for the AMI-based interlayer correlation ($r$), resulting in non-monotonic divergence values with respect to $d$, which is consistent with our hypothesis of parameters' interdependence. Nevertheless, the lowest divergence is attained for $d=2$, which coincides with the design choice recommended by the authors of the \mABCD\ model.

An additional, perhaps less anticipated, effect of using a high $d$ value is the increased divergence in edge correlation ($R$). The growing misalignment of communities between layers, when comparing the original network to the digital twin, creates unfavourable conditions for efficient link rewiring (phase 6 of the \mABCD\ process). It should also be noted that relatively high values of $\gamma$ and $\beta$ can stem from the non necessarily power-law-like distribution of degrees and communities in \textit{Freebase}. A potential improvement for that could be injecting sequences of them directly into \mABCD\ and, by that, to pass over some of its phases.

\subsection{Experiment 2 -- Configuration Retrieval Methods}

The objective of the second experiment was to evaluate various \mABCD\ configuration retrieval methods within the Bayesian framework outlined in Sec.~\ref{sec:extracting_parameters}. Specifically, we are interested in the following research question. Does increasing the fidelity of the produced twin in one measured dimension trade off for a decrease in other scores (\emph{Open Problem~6})? To make a step toward this direction, we investigated whether a specific Bayesian tuning configuration exists that outperforms the others across all divergence dimensions.

Based on the considerations, four configuration retrieval runs were executed with the following specifications:
\begin{enumerate}
    \item decision variable $r$ with loss based on $\mathcal{D}_{r} ( \hat{G}, G )$;
    \item decision variables $r$ and $\tau$ with loss based on $\mathcal{D}_{r} ( \hat{G}, G )$;
    \item decision variables $r$ and $\tau$ with loss based on $\mathcal{D}_{\tau} ( \hat{G}, G )$;
    \item decision variables $r$ and $\tau$ with loss based on $(\mathcal{D}_{r} ( \hat{G}, G ) + \mathcal{D}_{\tau} ( \hat{G}, G ) ) / 2$.
\end{enumerate}

\begin{figure}[ht!]
    \centering
    \vspace{-2em}
    \includegraphics[scale=0.51]{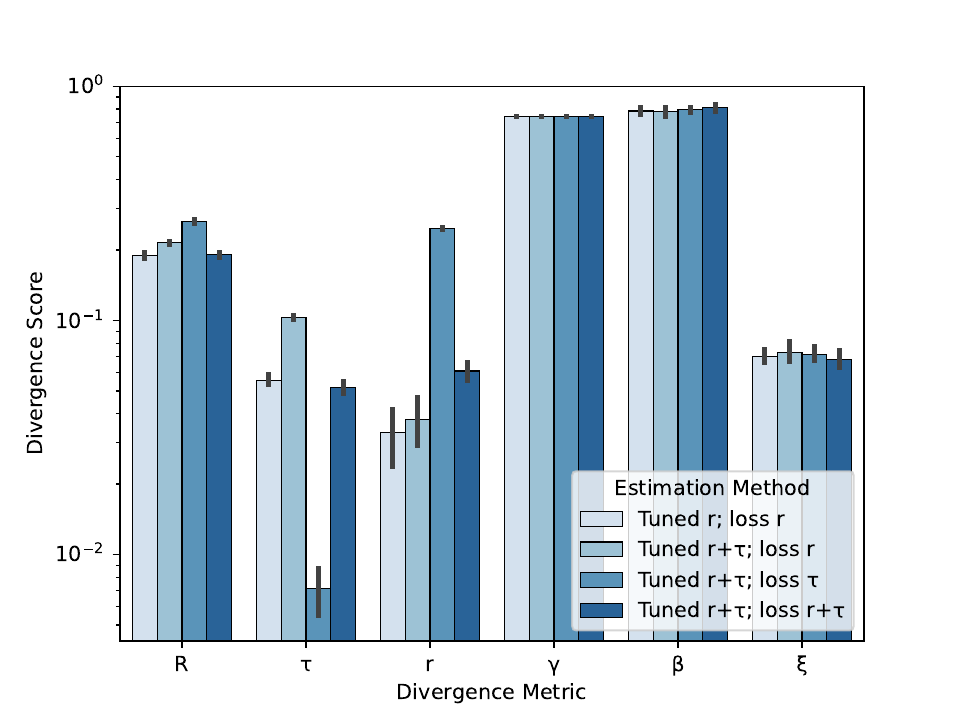}
    \caption{Divergence scores for four configuration estimation methods; fixed $d=2$.} 
    \label{fig:experiment_2}
\end{figure}

Based on the outcomes of Experiment 1, we set $d=2$ for all configuration retrieval specifications. Subsequently, ten digital twins were generated for each set of derived \mABCD~parameters, and the divergence metrics were calculated (Fig.~\ref{fig:experiment_2}).

The resulting scores for $\gamma, \beta,$ and $\xi$ once again validate correct implementation of the configuration retrieval: their minimal variation is attributed to the stochastic nature of \mABCD\ and the community extraction algorithm. More significant findings, consistent with expectations, are observed in the divergence metrics $\mathcal{D}_{\tau} ( \hat{G}, G )$ and $\mathcal{D}_{r} ( \hat{G}, G )$. Both scores are minimised when the optimisation technique targets the corresponding loss function (either $\mathcal{D}_{\tau} ( \hat{G}, G )$ or $\mathcal{D}_{r} ( \hat{G}, G )$). However, a trade-off is evident in the form of increased divergence for the alternative metric. This is particularly pronounced for the $\tau$-based loss method, which achieves a divergence score of $0.0071$, compared to an average of $0.0701$ across other specifications. Conversely, for the $\mathcal{D}_{r} ( \hat{G}, G )$ criterion, the $\tau$-based loss yields $0.2469$, indicating suboptimal performance compared to the average of $0.0438$ for other losses. Interestingly, expanding the set of decision variables from a single $r$ to include both $r$ and $\tau$ (while maintaining $\mathcal{D}_{r} ( \hat{G}, G )$ loss) does not decrease the $\mathcal{D}_{r} ( \hat{G}, G )$ score, but notably exacerbates the $\mathcal{D}_{\tau} ( \hat{G}, G )$. This phenomenon may be attributed to the curse of dimensionality and insufficient exploration of the parameter space. Variations in the $R$ score are difficult to attribute definitively; however, given their small magnitude, it can be concluded that the specific choice of decision variables and loss functions has a negligible impact on edge correlation fidelity.

\begin{figure}[ht!]
    \centering
    \includegraphics[scale=0.4]{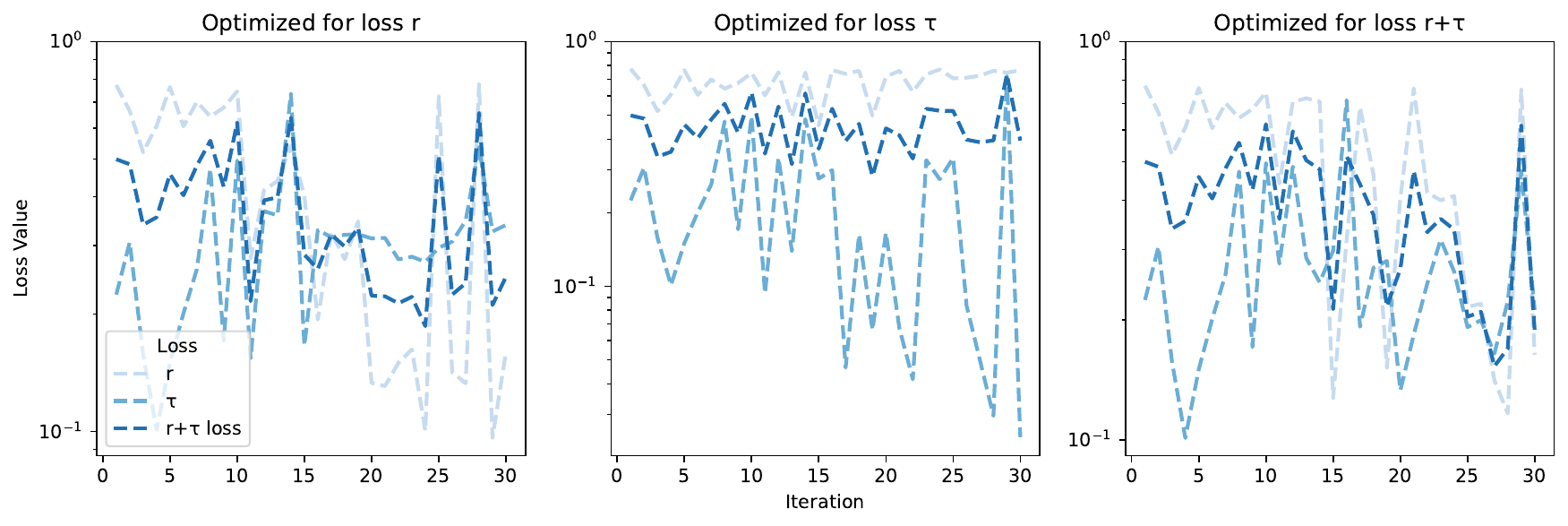}
    \caption{$\mathcal{D}_{\tau} ( \hat{G}, G )$, $\mathcal{D}_{r} ( \hat{G}, G )$, and $(\mathcal{D}_{r} ( \hat{G}, G )+\mathcal{D}_{\tau} ( \hat{G}, G )) /2$ scores as functions of the optimisation step in finding the best matching $r$ or $\tau$.} 
    \label{fig:experiment_2_loss_current}
\end{figure}

To complement the analysis of Experiment 2, we present the loss trajectories for the optimisation model utilising both $r$ and $\tau$ as decision variables. Fig.~\ref{fig:experiment_2_loss_current} illustrates the instantaneous loss calculated at each iteration of the search. It should be noted that we report $\mathcal{D}_{\tau} ( \hat{G}, G )$, $\mathcal{D}_{r} ( \hat{G}, G )$, and $(\mathcal{D}_{r} ( \hat{G}, G )+\mathcal{D}_{\tau} ( \hat{G}, G ))/2$ for each loss used as the objective function in the optimisation problem. Consistent reductions in the loss associated with the primary objective are observable, highlighting the efficacy of the Bayesian optimisation. Conversely, the trajectories for the non-target criterion predominantly oscillate around the average, indicating that any reduction in the alternative loss occurs incidentally rather than through directed optimisation.

%%%%%%%%%%%%%%%%%%%%%%%%%%%%%%%%%%%%%%%%%%%%%%%%%%%%%%%%%%%
\section{Conclusions\label{sec:conclusions}}
%%%%%%%%%%%%%%%%%%%%%%%%%%%%%%%%%%%%%%%%%%%%%%%%%%%%%%%%%%%

In this work, we discussed a problem of configuration retrieval for multilayer networks with the synthetic network generator --- \mABCD. Addressing that issue is vital for twinning complex networked systems and opens new possibilities for graph-based data augmentation in machine learning applications.

The main contribution of this study is the proposal of a statistical network inference method from empirical multilayer datasets. We also introduce several divergence measures for assessing the quality of parameter estimation. The method is evaluated on a real-world network representing cooperation within the movie industry. The results provide preliminary evidence supporting the validity of the proposed approach; however, a key finding is that independent parameter estimation techniques encounter intrinsic accuracy limits, as the configuration parameters collectively determine the structural properties of networks generated by \mABCD. This observation naturally motivates the development of joint prediction methods as a promising future direction.

Nevertheless, as this work is rather a perspective paper, we left readers with six \textit{O}pen \textit{P}roblems that can serve as a roadmap for deeper investigation and which (we believe) will open a fruitful discussion.

\begin{enumerate}[label={\textit{OP-\arabic*}}, leftmargin=*]
    \item assessing whether utilising deep embedding methods for joint-parameter estimation is feasible;
    \item understanding the influence of the hyper-parameter $d$ and selecting its optimal value;
    \item handling networks with degree sequences that do not follow the power-law by injecting them into \mABCD;
    \item handling networks with community size sequences that do not follow the power-law by injecting them into \mABCD;
    \item providing a cumulative divergence score, a scalar value which encapsulates all structural discrepancies between the original network and the digital twin;
    \item assess if increasing the fidelity of the produced twin with respect to one parameter trades off with a decrease in other scores.
\end{enumerate}

\begin{credits}

\subsubsection{\ackname}
This research was partially supported by: (1) EU under the Horizon Europe, grant no. 101086321 OMINO; (2) Polish Ministry of Science and Higher Education, International Projects Co-Funded programme; (3) National Science Centre, Poland, grant no. 2022/45/B/ST6/04145; (4) Polish National Agency for Academic Exchange, Strategic Partnerships programme, grant no. BPI/PST/2024/1/00129/U/00001; (5) Wrocław University of Science and Technology, Academia Profesorum Iuniorum programme. Views and opinions expressed are, however, those of the authors only and do not necessarily reflect those of the founding agencies. 

\subsubsection{\discintname} The authors have no competing interests to declare that are relevant to the content of this article.

\subsubsection{Code and Data} The methods presented in this manuscript have been implemented in Python. The source code and installation instructions are available at: \url{https://github.com/network-science-lab/mabcd-for-digital-twins}.

\end{credits}

\bibliographystyle{splncs04}
\bibliography{references}

\end{document}